\documentclass[sigconf]{acmart}

\usepackage{caption}
\usepackage{graphicx}
\usepackage{xspace}
\usepackage{cleveref}
\usepackage{lipsum}
\usepackage[normalem]{ulem}
\usepackage{enumitem}
\usepackage{geometry}
\usepackage{amsmath}
\setlist[itemize]{noitemsep, topsep=0pt}
\usepackage[utf8]{inputenc}
\usepackage[T1]{fontenc}
\usepackage{adjustbox}
\usepackage{subfigure}
\usepackage{multirow}
\usepackage[normalem]{ulem}

\usepackage{soul}
\usepackage{hyperref}
\hypersetup{
    colorlinks=true,
    linkcolor=blue,
    filecolor=magenta,
    urlcolor=cyan,
}

\usepackage{siunitx}
\usepackage{tikz}

\makeatletter
\DeclareRobustCommand\onedot{\futurelet\@let@token\@onedot}
\def\@onedot{\ifx\@let@token.\else.\null\fi\xspace}

\makeatother

\AtBeginDocument{%
  \providecommand\BibTeX{{%
    \normalfont B\kern-0.5em{\scshape i\kern-0.25em b}\kern-0.8em\TeX}}}

\begin{document}
\title[]{Spatial Discretization for Fine-Grain Zone Checks with STARKs}



\author{Sungmin Lee}
\email{i.am.sungmin.lee@yonsei.ac.kr}
\affiliation{%
  \institution{Yonsei University}
  \country{}
}

\author{Kichang Lee}
\email{kichang.lee@yonsei.ac.kr}
\affiliation{%
  \institution{Yonsei University}
  \country{}
}

\author{Gyeongmin Han}
\email{sd061123@yonsei.ac.kr}
\affiliation{%
  \institution{Yonsei University}
  \country{}
}

\author{JeongGil Ko}
\email{jeonggil.ko@yonsei.ac.kr}
\affiliation{%
  \institution{Yonsei University}
  \country{}
}


\begin{abstract}

Many location-based services rely on a point-in-polygon test (\textit{PiP}), checking whether a point or a trajectory lies inside a geographic zone.
Since geometric operations are expensive in zero-knowledge proofs, privately performing the \textit{PiP} test is challenging.
In this paper, we answer the research questions of how different ways of encoding zones affect accuracy and proof cost by exploiting grid-based lookup tables under a fixed STARK execution model.
Beyond a Boolean grid-based baseline that marks cells as in- or outside, we explore a distance-aware encoding approach that stores how far each cell is from a zone boundary and uses interpolation to reason within a cell.
Our experiments on real-world data demonstrate that the proposed distance-aware approach achieves higher accuracy on coarse grids (max. $60\%p$ accuracy gain) with only a moderate verification overhead (approximately $1.4\times$), making zone encoding the key lever for efficient zero-knowledge spatial checks.
\end{abstract}
\settopmatter{printfolios=true} 
\settopmatter{printacmref=false} 
\renewcommand\footnotetextcopyrightpermission[1]{} 
\maketitle

\vspace{-2ex}
\section{Introduction}

Time-series trajectory data generated by mobile and cyber-physical systems underpins various location-based classification services, including geofencing and region-based regulation~\cite{demontjoye2013unique, Garzon2015Geofencing, lee2025gmt}.
At the core of these applications lies a fundamental question: \emph{does a given location or trajectory lie inside a specified geographic zone?}
This \emph{zone checking} (i.e., \textit{point-in-polygon} (\textit{PiP})~\cite{HormannAgathos2001PiP}) problem is challenging in practice, as real-world zones exhibit narrow roads, irregular boundaries, and overlapping jurisdictions, where even small geometric errors can flip compliance outcomes.

Zero-knowledge proofs (ZKPs) have recently been explored to make such compliance audit verifiable without trusting the computing party~\cite{fiege1987zero, boo2021litezkp, Lin2021TrajectoryZKBlockchain}.
Unfortunately, spatial predicates are poorly matched to arithmetic proof systems, which introduce challenges in applying ZKPs to the zone checking problem.
Standard geometric operations—such as \textit{PiP} tests—rely on branching and comparisons, which translate inefficiently into the arithmetic constraints required by SNARKs~\cite{Groth2016Groth16} and STARKs~\cite{BenSasson2018STARK}, leading to large and costly proofs.

To mitigate this discrepancy, previous works replaced expensive computation with lookup tables over precomputed values~\cite{gabizon2020plookup}.
This approach shifts proof-time execution to simple indexing and arithmetic, which aligns well with modern ZKP cost models~\cite{habock2022multivariate}.
Applied to spatial predicates, this approach naturally yields \emph{lookup-based discretization}, where the 2D plane is discretized into grid cells and spatial information is stored per cell.
At proof time, a location is mapped to a cell index, and zone membership is evaluated via bounded lookups rather than in-circuit geometry.

Discretization, however, introduces inherent trade-offs: coarse grids reduce proof cost but discard sub-cell geometric detail, causing errors near boundaries, while finer grids improve accuracy at the expense of higher proving costs~\cite{HormannAgathos2001PiP}.
Thus, the choice of discretization strategy and resolution is crucial for \textit{PiP} test, since it fundamentally shapes both spatial accuracy and proof efficiency.

In this paper, we systematically study these accuracy-cost trade-offs under STARK by addressing the following research questions.

\vspace{0.04in}
\noindent$\bullet$\textbf{RQ1:} How do different lookup-based discretization strategies compare in terms of spatial accuracy and zero-knowledge proof cost?

\vspace{0.04in}

\noindent$\bullet$\textbf{RQ2:} How does grid resolution affect this trade-off, and what resolution ranges are practically effective?
\vspace{0.04in}

To answer the aforementioned research questions, we examine three representative grid-based discretization strategies:
(i) \emph{Center-point} discretization assigns zone membership based on each cell’s center, offering minimal overhead but accuracy limited to the grid scale.
(ii) A \emph{voting-based} variant aggregates zone coverage within a cell, reducing some boundary sensitivity but often introducing staircase artifacts.
(iii) Beyond these boolean baselines, we consider a \emph{distance-aware} encoding that stores signed distance values~\cite{hart1996sphere} at grid vertices and reconstructs sub-cell geometry via bilinear interpolation~\cite{smith1981bilinear}.
Because interpolation uses only additions and multiplications, it is well-suited to arithmetic-based ZKP systems while preserving boundary proximity information.


Through an end-to-end evaluation on real-world geographic zones~\cite{OpenStreetMap} with a full STARK implementation, we demonstrate that the discretization is not merely a matter of grid size.
While increasing resolution quickly incurs diminishing accuracy gains, richer distance-aware encodings fundamentally improve the accuracy--cost trade-off, achieving large accuracy gains even on coarse grids (up to $60\%p$ at $16\times16$ resolution) with only $1.4\times$ additional verification latency.

Our contributions are threefold.
(i) We formalize lookup-based spatial predicate evaluation under a unified STARK execution model and empirically compare boolean and distance-aware discretization strategies (\textbf{RQ1}).
(ii) We quantify how grid resolution governs the accuracy-cost trade-off and identify practically effective resolution regimes (\textbf{RQ2}).
(iii) We provide an end-to-end evaluation on real-world geographic data, demonstrating how discretization choices translate into concrete proving and verification costs.
\section{Background and Related Work}

\subsection{STARK and Lookup-Based Computation}

STARKs encode computation as algebraic constraints over finite fields,
where execution is represented as a trace and correctness is enforced
through low-degree polynomial relations~\cite{BenSasson2018STARK}.
In this model, proof cost is dominated by the length and width of the
execution trace and by lookup and consistency verification, while
verification cost depends only weakly on trace size~\cite{Thaler2018Proofs}.

Geometric predicates are fundamentally mismatched to this execution
model.
Standard geometric operations rely on branching and comparisons, which
translate inefficiently into arithmetic constraints~\cite{Groth2016Groth16}.
To mitigate this mismatch, modern STARK designs rely on lookup arguments,
which replace in-circuit computation with authenticated access to
precomputed tables~\cite{gabizon2020plookup}.
Under this paradigm, complex logic is executed during preprocessing, and
the proof verifies only index correctness and table membership.

Lookup-based computation has therefore become a central design pattern in
STARK-friendly systems~\cite{gabizon2020plookup}.
By fixing the execution structure and shifting complexity into public
tables, lookup arguments yield predictable proof cost.
This work adopts the same principle: we fix the STARK execution model and
vary only the spatial information stored in lookup tables, isolating how
different encodings affect accuracy and proof cost.

\subsection{Spatial Predicates and Discretization under Zero Knowledge}

Spatial predicates such as zone membership are a core building block of
location-based applications, including geofencing, regulatory
compliance, and trajectory analysis~\cite{demontjoye2013unique, Garzon2015Geofencing}.
Direct geometric evaluation is precise but poorly suited to arithmetic
proof systems, motivating discretization-based alternatives.

Discretization replaces continuous space with a finite grid and stores
precomputed spatial information per cell.
At proof time, a query location is mapped to a cell index, and zone
membership is derived from table lookups rather than explicit
geometry~\cite{gabizon2020plookup}.
Existing approaches predominantly rely on boolean encodings, such as
center-point tests or area-based aggregation, which discard sub-cell
geometry and introduce boundary errors that diminish only with
increasing grid resolution~\cite{HormannAgathos2001PiP}.

Signed distance representations~\cite{hart1996sphere} provide a richer alternative by encoding
boundary proximity rather than binary membership~\cite{HormannAgathos2001PiP}.
This paper shows that, when combined with arithmetic-native
interpolation, distance-aware encodings fit naturally within a
lookup-based STARK model and improve the accuracy-cost trade-off without
changing the underlying proof structure.
\section{Methodology}
\label{sec:methodology}

This section presents our methodology for studying lookup-based discretization of spatial predicates under a fixed STARK execution model.
Our focus is deliberately narrow: we keep the proof structure and arithmetization unchanged, and vary only the \emph{spatial information stored in public lookup tables}.
This isolates how different discretizations affect approximation accuracy and proof cost.
Concretely, we evaluate multiple discretizations using a single, reusable STARK AIR template for per-zone membership queries.

\subsection{Problem Setting}

We consider a public axis-aligned bounding box $B = [x_{\min},x_{\max}]\times[y_{\min},y_{\max}]$ and a public set of polygonal zones $\mathcal{Z}=\{z_1,\dots,z_m\}$ defined within $B$.
A prover holds a private query location $(x,y)\in B$ (or a sequence of
such points in the trajectory setting) and must convince a verifier of
the correctness of zone membership predicates
$\mathsf{inside}(x,y,z_i)\in\{0,1\}$ for a specified subset of zones.

We assume a standard STARK setting in which the prover’s inputs remain
hidden except for the final membership outputs.
Our goal is not to analyze zero-knowledge guarantees, but to study how
different spatial table encodings behave under the same STARK-friendly
computation.

\subsection{Lookup-Based Discretization (RQ1)}
\label{subsec:lookup_discretization}

Lookup-based discretization replaces in-circuit geometric computation
with authenticated access to precomputed spatial tables.
Direct predicates such as point-in-polygon tests rely on branching and
comparisons that are expensive to arithmetize, leading to large STARK
execution traces.
By contrast, lookup-based discretization shifts this cost to
preprocessing, and the proof reasons only about table indices and
retrieved values.

At proof time, a query location $(x,y)$ is mapped to a grid cell $(i,j)$,
table entries are retrieved via lookup, and membership is derived using
lightweight arithmetic.
Under this model, proof cost is governed by table size, lookup
verification, and the arithmetic needed to interpret the stored payload.
This makes the choice of \emph{what information to encode in the lookup
tables} the central design decision studied in \textbf{RQ1}.

\subsubsection{Grid-Based Lookup Model}

We discretize the public bounding box
$B$ into a uniform
$r\times r$ grid.
Rather than maintaining separate tables per zone, we conceptually
represent all zones using a unified lookup table $T : \mathcal{Z} \times \{0,\dots,r-1\}^2 \rightarrow \mathbb{F}$,
which maps each zone and grid cell to a stored payload.

All discretization strategies share the same grid resolution, indexing
scheme, and lookup interface, and differ only in the payload associated
with each $(z,i,j)$ entry.
This formulation isolates discretization to the table contents and
ensures that different encodings are evaluated under an identical STARK
execution model.

\begin{figure}
    \centering
    \includegraphics[width=.9\linewidth]{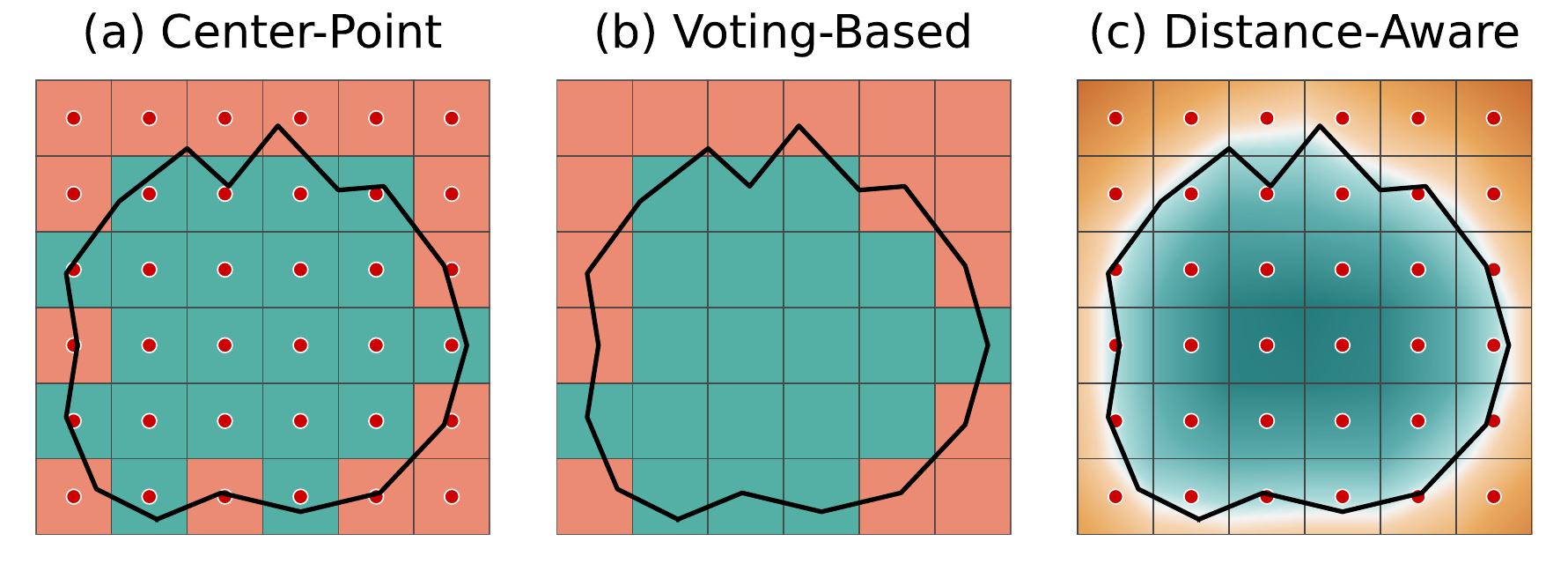}
    \vspace{-3ex}
    \caption{Comparison of lookup-based discretization strategies. The color
gradient in panel (c) visualizes continuous boundary proximity recovered via bilinear interpolation.}
    \vspace{-3ex}
    \label{fig:overview}
\end{figure}

\subsubsection{Center-Point Discretization (Boolean)}

The center-point encoding stores a single bit per cell indicating
whether the cell center lies inside the zone (Fig.~\ref{fig:overview}(a)): $T[z,i,j] = \mathsf{inside}(c_x,c_y,z)$,
where $(c_x,c_y)$ denotes the center of cell $(i,j)$.
Membership is given directly by the looked-up value, requiring no
additional arithmetic beyond boolean checks.

This encoding has minimal table size and trace width, making it the
cheapest option in proof cost.
However, it discards all sub-grid geometry and introduces boundary
errors whenever a cell straddles a zone boundary.
As a result, its accuracy is fundamentally limited by the grid
resolution.

\subsubsection{Voting-Based Discretization (Boolean)}

Voting-based discretization aggregates geometric coverage within each
cell (Fig.~\ref{fig:overview}(b)).
Each cell stores a boolean indicator whether the zone occupies
more than a fixed fraction of the cell area, determined during
preprocessing:
\[
T[z,i,j] = \mathbf{1}\!\left[
\frac{\mathrm{area}(\mathrm{cell}(i,j)\cap z)}
     {\mathrm{area}(\mathrm{cell}(i,j))} \ge \tau
\right],
\]
where $\tau$ is a fixed threshold (e.g., $\tau=0.5$).

Compared to center-point encoding, voting can reduce errors caused by
unfortunate boundary alignment.
However, it still collapses partial coverage into a hard decision and
eliminates information about how close a query lies to the true
boundary.
This often leads to staircase artifacts and unstable behavior for thin
or fragmented regions, while retaining the same boolean semantics and
limited expressiveness.

\subsubsection{Distance-Aware Discretization (SDF + Interpolation)}

To overcome the loss of sub-cell geometric information inherent in
boolean discretization, we propose a distance-aware encoding that
supports \emph{sub-grid spatial resolution} while remaining compatible
with STARK-style arithmetic execution (Fig.~\ref{fig:overview}(c)).
The key idea is to store boundary proximity information and recover
finer spatial detail through simple arithmetic, rather than by
increasing grid resolution or performing explicit geometric tests.

Instead of storing binary membership, we associate each grid vertex with
a signed distance to the zone boundary.
For a point $p$ and zone $z$, the signed distance function is defined as
\[
\mathsf{SDF}(p,z)=
\begin{cases}
-d(p,\partial z) & \text{if } p\in z,\\
+d(p,\partial z) & \text{otherwise},
\end{cases}
\]
where $d(\cdot,\partial z)$ denotes Euclidean distance.
Distances are represented using fixed-point arithmetic (scale
$S=2^{32}$), allowing them to be manipulated using field operations.

At query time, we reconstruct a \emph{sub-grid estimate} of the signed
distance by interpolating SDF values stored at the four grid vertices
surrounding the query point.
Let $T[z,i,j]$ denote the signed distance value associated with zone $z$
at grid vertex $(i,j)$.
This enables continuous, within-cell reasoning from discrete lookup
tables.

We use bilinear interpolation~\cite{smith1981bilinear} because it can be expressed using only
additions and multiplications and is therefore arithmetic-native and
well suited to STARK execution models.
The interpolation weights are determined by the relative position of
$(x,y)$ within the grid cell and are normalized to lie in $[0,1]$, so the
resulting value is a weighted average of the four neighboring SDF
values.

Because this interpolation forms a convex combination, the estimated
distance varies smoothly within each cell and provides a continuous
sub-grid estimate of boundary proximity.
Zone membership is determined by the sign of the interpolated value:
negative distances indicate points inside the zone, while positive
values indicate points outside.

Overall, combining signed distances with interpolation preserves
boundary information discarded by boolean tables, allowing coarse grids
to approximate fine-grained geometry.
The additional cost is limited to a small number of extra lookups and
arithmetic operations, resulting in a constant increase in trace width
and verification time under the fixed AIR.

\section{Evaluation}
\label{sec:evaluation}

We evaluate grid-based spatial discretization strategies to answer two research
questions: \textbf{RQ1}, how different discretization strategies compare in
terms of spatial accuracy and zero-knowledge proof cost, and \textbf{RQ2}, how
grid resolution itself shapes the accuracy-cost trade-off.

\subsection{Experimental Setup}
\label{sec:exp-setup}

\subsubsection{Datasets}

\begin{figure}
    \centering
    \includegraphics[width=.9\linewidth]{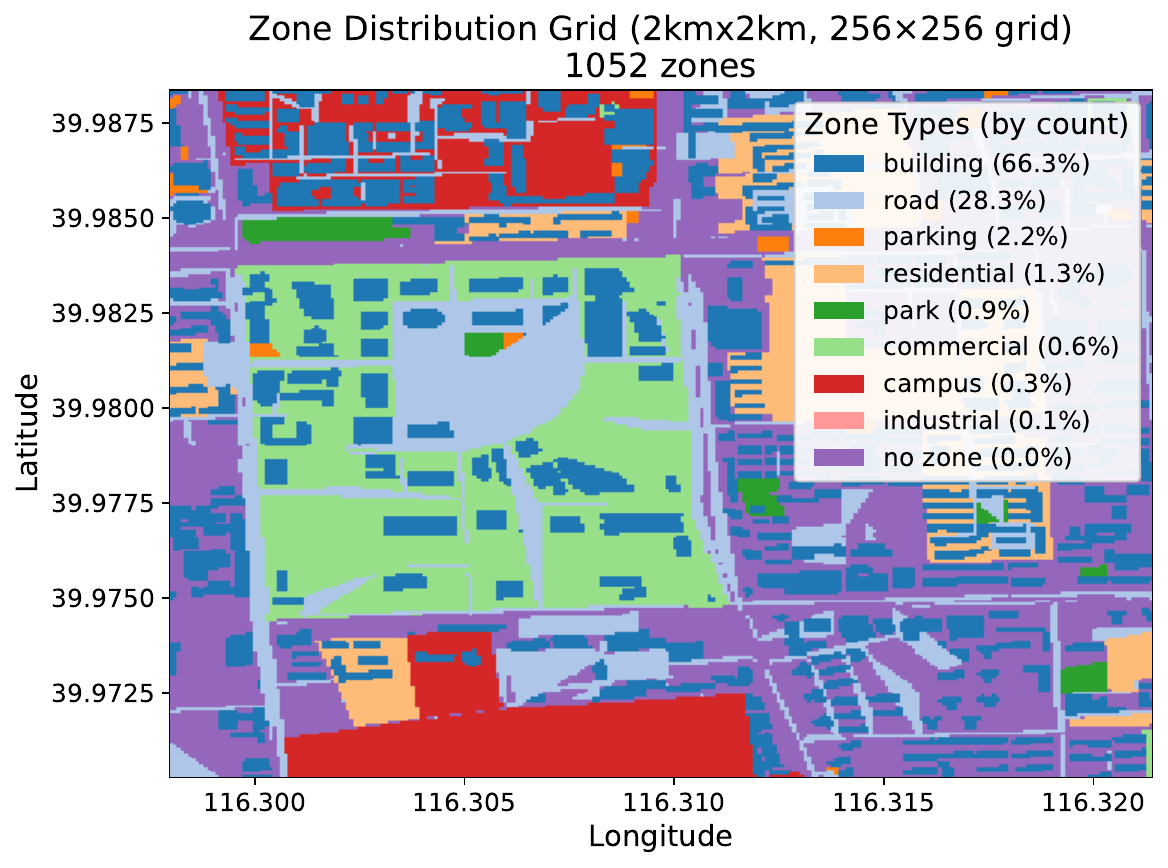}
    \vspace{-2ex}
    \caption{Spatial distribution of semantic zone categories of our dataset.
Colors indicate representative zone types for visualization, while zones may
overlap in practice. The figure highlights heterogeneous geometries, including compact buildings and narrow road segments.}
    \vspace{-4ex}
    \label{fig:zone_dist}
\end{figure}

Real-world spatial zones are extracted from OpenStreetMap~\cite{OpenStreetMap}.
We consider nine semantic zone categories, including buildings, parks,
residential areas, roads, and campuses, as illustrated in
Fig.~\ref{fig:zone_dist}.
Because a single location may belong to multiple zones simultaneously, all
evaluations are conducted in a multi-zone setting; figures visualize zones by
their representative categories for clarity.
Experiments focus on a $2\,\text{km} \times 2\,\text{km}$ urban region in Beijing,
which contains 1{,}052 zone polygons with diverse geometric complexity.

To evaluate discretization effects, the region is partitioned into an
$r \times r$ grid with resolution
$r \in \{2, 3, 4, 8, 16, 32, 64, 128, 256\}$.
This range spans coarse discretizations ($2\times2$, cell size
$\approx1$\,km) to finer grids ($256\times256$, cell size $\approx8$\,m),
capturing the transition from severe boundary approximation to substantially
improved granularity.

\subsubsection{Metrics}

We measure spatial accuracy and zero-knowledge proof cost.
For each grid resolution, spatial accuracy is computed using 4{,}000 randomly
sampled query points within the bounding box.
Since a query point may belong to multiple zones simultaneously, accuracy is
evaluated in a multi-zone setting: a prediction is considered correct only if
the discretized membership decisions match the exact geometric containment
results for all queried zones.
This metric captures discretization-induced errors that arise from boundary
approximation and zone overlap.

Proof cost is measured using wall-clock proving time and verification time.
Each measurement is repeated three times, and we report the average.
Proving time includes full execution trace construction and STARK proof
generation, while verification time measures end-to-end proof validation.

\subsubsection{Hardware and Implementation}

All experiments were conducted on a single machine equipped with an AMD Ryzen~9 9950X3D CPU (16 cores, 32 threads at 4.3\,GHz) and 60\,GB of DDR memory. The implementation is written in Python~3.12, using NumPy for array operations and Shapely for geometric computations. The STARK proof system is implemented as a research-oriented prototype using standard cryptographic primitives (e.g., SHA-256) and NumPy-based arithmetic, without performance optimizations.

\begin{figure}
    \centering
    \includegraphics[width=1\linewidth]{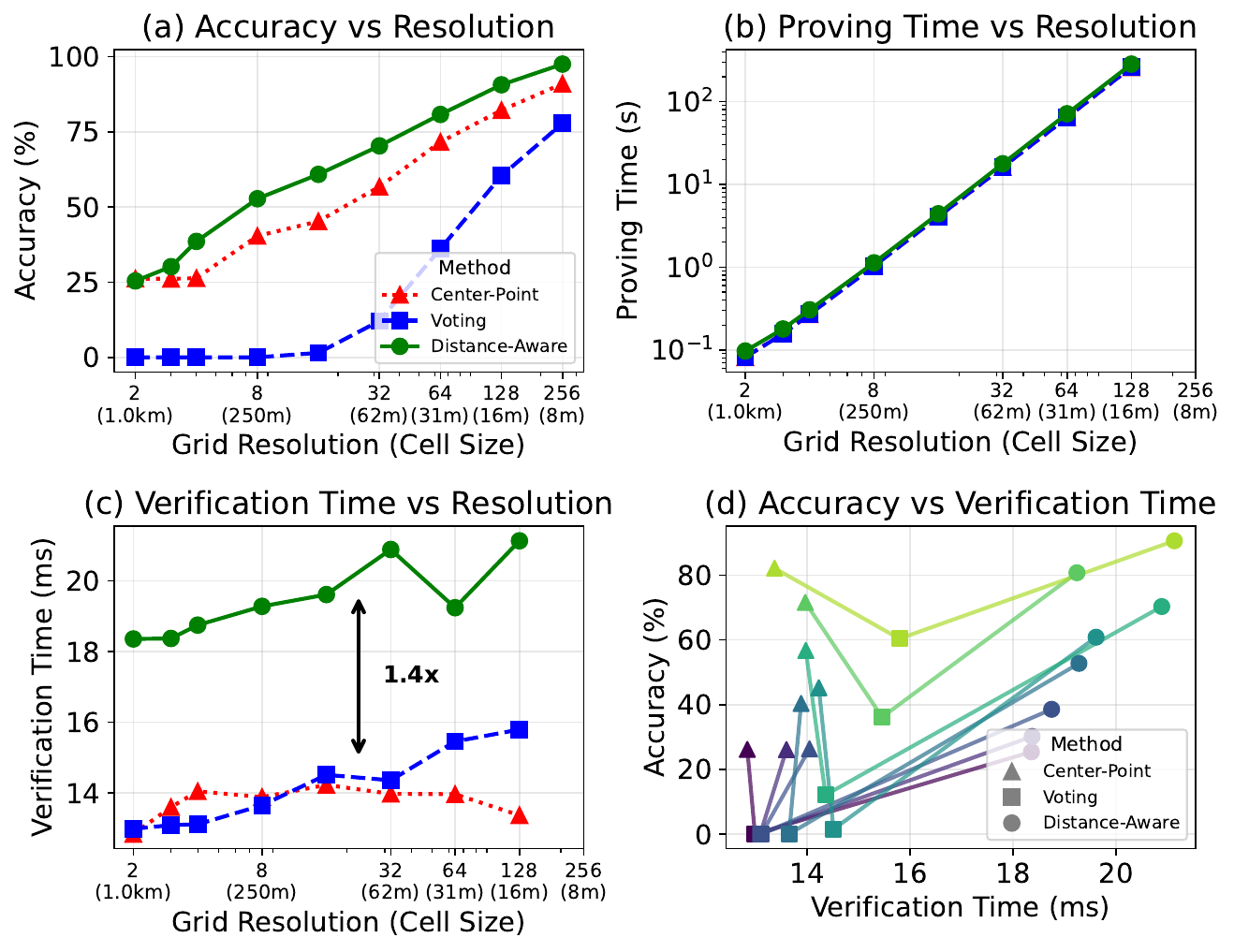}
    \vspace{-5ex}
    \caption{Accuracy and proof-cost trends across grid resolutions and
discretization strategies.
For panel (a), (b), and (c), lines and markers correspond to center-point (red triangles), voting-based (blue squares), and distance-aware (green circles) discretization.
In panel (d), color encodes grid resolution: darker purple indicates coarser
grids (from $2\times2$), while brighter green corresponds to finer grids (up to
$128\times128$).}

    \vspace{-4ex}
    \label{fig:RQ1}
\end{figure}

\subsection{RQ1: Comparing Discretization Strategies}
\label{sec:rq1}

We first address \textbf{RQ1}: how grid-based discretization strategies compare
in terms of spatial accuracy and zero-knowledge proof cost.
Fig.~\ref{fig:RQ1} summarizes the relationship between accuracy, proving time,
verification time, and grid resolution.

Fig.~\ref{fig:RQ1}(a) shows that spatial accuracy improves monotonically with
increasing resolution for all methods, reflecting reduced discretization error.
However, convergence behavior differs substantially.
Center-point discretization converges slowest, while voting-based aggregation
exhibits irregular, alignment-sensitive gains. Distance-aware encoding consistently outperforms boolean baselines at coarse and
intermediate resolutions.
At very coarse grids ($8\times8$), distance-aware encoding already achieves
around $50\%$ accuracy while voting remains near zero.
The gap peaks at $16\times16$ resolution, where
distance-aware encoding reaches about $61\%$ accuracy compared to roughly $1\%$
for voting, yielding an absolute gain close to $60\%p$.
Even at $32\times32$, a large gap persists (about $70\%$ vs.\ $12\%$), before the
advantage gradually diminishes at finer resolutions as voting begins to catch up.
By storing signed distance values at grid vertices and interpolating within
cells, distance-aware encoding preserves boundary proximity information
discarded by boolean discretization, enabling smooth within-cell decisions and
higher accuracy at substantially lower resolutions.

Verification cost further highlights structural differences.
As shown in Fig.~\ref{fig:RQ1}(c), verification time is nearly constant across
resolutions, reflecting the resolution-insensitive nature of STARK verification.
Center-point and voting-based methods incur nearly identical costs, while
distance-aware encoding introduces a modest but consistent overhead of
approximately $1.4\times$.
This overhead arises from constant-factor effects rather than asymptotic scaling.
STARK verification performs a fixed number of Merkle path checks and FRI queries,
while distance-aware encoding additionally verifies multiple table lookups and
interpolation constraints.
These operations increase verifier workload by a constant factor, yielding
higher but resolution-invariant verification time.

Fig.~\ref{fig:RQ1}(d) visualizes the accuracy-verification trade-off.
Center-point and distance-aware encodings form the upper envelope, while
voting-based aggregation is consistently dominated.
Despite similar verification cost to center-point discretization, voting
achieves lower accuracy, creating a concave dip in the trade-off
space.

Overall, discretization strategy reshapes the accuracy-cost frontier through
constant-factor effects.
Boolean methods offer minimal verification overhead but differ in geometric
fidelity, whereas distance-aware encoding trades a small, predictable overhead
for substantially improved accuracy at coarse resolutions.

\subsection{RQ2: Resolution Effects and Prover-Side Cost Scaling}
\label{sec:rq2}

We next investigate the effect of grid resolution on the accuracy-cost trade-off.
As shown in Fig.~\ref{fig:RQ1}(a), accuracy improves with resolution for all
strategies but exhibits diminishing returns beyond moderate grid sizes.
For distance-aware encoding, gains plateau around $32\times32$, indicating that
most boundary structure is recovered before very fine discretization.

In contrast, proving time grows rapidly with resolution
(Fig.~\ref{fig:RQ1}(b)), consistent with superlinear scaling.
This follows from the STARK prover cost model: higher resolutions increase trace
length, FFT- and IFFT-based commitments, Merkle tree construction, FRI folding,
and lookup consistency checks.

Fig.~\ref{fig:RQ1}(d) further shows that resolution determines when
method-level trade-offs emerge.
At very coarse resolutions, all strategies cluster tightly and discretization
error dominates.
Beyond roughly $32\times32$, accuracy gains become visible and differences across
methods meaningfully shape the accuracy-verification frontier.

These results suggest that resolution selection precedes strategy selection.
Once a resolution escapes the coarse-discretization regime, discretization
strategy determines the trade-off shape.
Distance-aware encoding is particularly effective in this regime, achieving
near-saturated accuracy with only a modest, resolution-independent verification
overhead.

\section{Limitation and Future Work}

Our study considers uniform grids and simple bilinear interpolation for
STARK compatibility; adaptive grids or higher-order interpolation may
further improve boundary fidelity at the cost of increased proof
complexity.
We evaluate point-wise zone membership only, and extending
distance-aware encodings to trajectory-level or temporally aggregated
predicates remains open.
Finally, while we assume a zero-knowledge execution setting, we do not
analyze privacy leakage or ZK-specific implementation optimizations,
which are left for future work.

\section{Conclusion}
We presented a systematic study of lookup-based spatial discretization
for zone membership under a unified STARK execution model.
By comparing boolean and distance-aware table encodings, we showed that
encoding boundary proximity via signed distances and arithmetic-native
interpolation fundamentally improves the accuracy-cost trade-off,
especially at coarse grid resolutions.
Our results highlight discretization strategy as a key design for
efficient zero-knowledge spatial compliance, beyond grid resolution
alone.


\bibliographystyle{ACM-Reference-Format}
\bibliography{reference, eis-lab}
\clearpage

\end{document}